\documentclass[a4paper,11pt]{article}
\usepackage{pos}
\usepackage{graphicx}
\usepackage{wrapfig}
\usepackage{natbib}

\newcommand{\sqrtsNN}{\sqrt{s_\mathrm{NN}}}
\newcommand{\GeV}{\mathrm{GeV}}

\newcommand{\mysection}[1]{
{\flushleft\emph{#1}}}

\title{The impact of phase transitions on dilepton production}

\author*[a,c,d]{Renan Góes-Hirayama}
\author[b,c,d]{Hannah Elfner}

\affiliation[a]{Department of Physics, McGill University,\\ 3600 University Street, Montreal, QC H3A 2T8, Canada}

\affiliation[b]{GSI Helmholtzzentrum f\"ur Schwerionenforschung,\\ Planckstr. 1, 64291 Darmstadt, Germany}

\affiliation[c]{Institute for Theoretical Physics, Goethe University,\\ Max-von-Laue-Strasse 1, 60438 Frankfurt am Main, Germany}

\affiliation[d]{Helmholtz Research Academy Hesse for FAIR (HFHF), GSI Helmholtz Center, Campus Frankfurt, Max-von-Laue-Straße 12, 60438 Frankfurt am Main, Germany}

\emailAdd{hirayama@itp.uni-frankfurt.de}

\abstract{We calculate dielectron yields at beam energies relevant for the CBM experiment at FAIR and the STAR Beam Energy Scan. The production of intermediate mass dielectrons is expected to increase in presence of a phase transition due to the latent heat. In this work, we use the SMASH+vHLLE hybrid model to quantify this enhancement in two scenarios: with a toy equation of state consisting of the Bag Model and Hadron Resonance Gas, separated by a Maxwell-constructed transition; and with a more realistic equation of state containing a critical endpoint, based on the 3D Ising universality class. We confirm that the dilepton yield increases when the system passes through a first-order phase transition, by about 7\% in the realistic scenario.}

\FullConference{13th International Conference on Hard and Electromagnetic Probes of High-Energy Nuclear Collisions (HardProbes2026)\\
21-26 June 2026\\
Nashville, Tennessee, USA\\}

\begin{document}
\maketitle

\section{Introduction}

Mapping the phase diagram of strongly-interacting matter at finite baryon density is one of the central goals of current heavy-ion research. At vanishing baryon chemical potential, lattice QCD shows that the transition from hadrons to a quark-gluon plasma (QGP) is a smooth crossover. At larger $\mu_B$, effective models suggest that it becomes first-order and ends in a critical endpoint (CEP). Locating the CEP and the first-order line is the aim of the STAR Beam Energy Scan program at RHIC and of the upcoming CBM experiment at FAIR  \cite{Bzdak:2019pkr}. 

Dileptons are emitted throughout the entire evolution and leave the medium unscathed, so their spectra carry information integrated over the whole space-time history of the collision. This makes them promising probes of a first-order phase transition: if the system goes through one, it must release or absorb latent heat while converting between phases, which lengthens the time spent in the hot region and could enhance dilepton production. Studies using coarse-grained transport simulations have proposed this connection in low beam energy collisions, and found that the yield of low mass dileptons could potentially double with a phase transition in the equation of state (EoS) \cite{Seck:2020qbx,Savchuk:2022aev}. In these studies, however, changing the EoS alters more than the transition itself; the confined and deconfined phases are modified as well, so the effect of the transition cannot be separated cleanly. In this work, we address this question focusing on the intermediate invariant mass range (IMR, $1.3<m_{ee}/\GeV<3$), which is dominated by thermal radiation from the earliest stages, with the QGP as a major contributor.

\section{Theoretical Framework}

\mysection{SMASH-vHLLE hybrid model.} To simulate and evolve the collisions explored in this work, we employ a hybrid chain that combines the hadronic transport approach SMASH \cite{SMASH:2016zqf} with the (3+1)D hydrodynamic code vHLLE \cite{Karpenko:2013wva}. The initial nucleons are sampled by SMASH with a Woods-Saxon distribution, and made to collide using cross sections that match experimental data \cite{Rosenkvist:2025tcl}. The produced hadrons interact and propagate further until they obey a fluidization criterion (detailed below), and then used to construct an initial condition (IC) for the hydrodynamic stage. The fluid evolution follows the Israel-Stewart equations of motion including shear and bulk viscosity, but charge diffusion is neglected. Here, we explore two types of EoS to close the system of equations. As the fluid expands and becomes dilute, the cells with low energy density undergo particlization using the Cooper-Frye formula, including off-equilibrium corrections \cite{Karpenko:2015xea}. The sampled particles are then returned to SMASH for an ``afterburner'' evolution until freezeout.

When the beam energy is sufficiently high, the Lorentz-contracted nuclei become flat and the midrapidity region is approximately boost invariant. In this situation,
the medium is thought to hydrodynamize at a constant hyperbolic time $\tau_0$. We fix $\tau_0$ by the nuclear passing time of the two nuclei, as done in previous works \cite{Karpenko:2015xea,Schafer:2021csj}. At lower beam energies, these conditions are not met, and there is no obvious kinematic region where to start the hydrodynamic evolution. Thus, we employ dynamical fluidization (DynFlu) \cite{Goes-Hirayama:2025nls}, where particles become fluid if the surrounding energy density surpasses a threshold $\varepsilon_\mathrm{th}$. In the case of fragmentation products, their proper time should exceed a fixed time $t_f$ in their rest frame, to prevent unphysical hotspots. Here, we set $\varepsilon_\mathrm{th}=0.35\ \mathrm{GeV/fm}^3$, and $t_f=0.2\ \mathrm{fm}$, with the remaining parameters following the Bayesian Analysis performed in \cite{Gotz:2025wnv}.

\mysection{Dilepton radiation.} In the off-equilibrium stages (IC and afterburner), SMASH uses the \emph{shining method}, where a resonance emits a perturbative dilepton every time step that it propagates, where each dilepton is associated to a weight proportional to the corresponding partial decay width, taken into account during analysis. The channels included are described in \cite{Staudenmaier:2017vtq}.

The fluid evolution in vHLLE uses thermal rates calculated in a many-body theory framework \cite{Rapp:1999us,vanHees:2007th}, which depend on the thermodynamic information extracted from each cell in the numerical grid, that is, the local temperature, total baryon density, and fraction of QGP. The total spectra is calculated by the library developed and available in \cite{Goes-Hirayama:2025lhs,Goes-Hirayama:2026ies}.

This analysis focuses on IMR dileptons, which are expected to be dominated by the QGP.  Since there are no resonant peaks in this region and the invariant mass spectrum is not shifted by radial flow, the slope of the spectrum can be used as a measure of the highest temperatures reached by the medium. As shown in \cite{Churchill:2023zkk,Goes-Hirayama:2025lhs}, the maximum temperatures reached by the medium are well correlated with  $T_\mathrm{eff}$ of a thermal spectrum, where we fit $
\dfrac{dN}{dm_{ee}} \propto m_{ee}^{3/2}\exp\left(-\dfrac{m_{ee}}{T_\mathrm{eff}}\right).$
    
\section{Results}

\mysection{Bag model + HRG.} In order to explore the sensitivity of dilepton production to a first-order phase transition, we construct a toy model that allows for a variable transition strength:
\begin{equation}\label{eq:Bag}
    P(T,\mu_B;B)=P_\mathrm{HRG}\Theta(T_c-T) + (P_\mathrm{partons}-B)\Theta(T-T_c).
\end{equation}
Here, the confined phase is described by $P_\mathrm{HRG}$, a hadron resonance gas including the SMASH degrees of freedom, $P_\mathrm{partons}$ is the pressure of a non-interacting ideal gas of quarks (two flavors) and gluons, and $B$ is the so-called Bag parameter. Increasing this parameter increases both the critical temperature $T_c$ -- determined by imposing continuity in the pressure \eqref{eq:Bag} -- and the latent heat, which is $\Delta \varepsilon=4B$ at $\mu_B=0$. The coexistence phase is determined via Maxwell construction, which automatically provides the fraction of QGP needed for the thermal rates. Equation \eqref{eq:Bag} is written as a function of $(T,\mu_B)$, but hydrodynamics requires a grid in $(\varepsilon,n_B)$. Such inversion --  for this EoS -- became numerically unreliable above $\mu_B>0.45\ \GeV$, so we limited its applicability to collisions of $\sqrtsNN\geq20\ \GeV$, avoiding the issue altogether. Because of this, we use an IC at constant $\tau_0$.

\begin{figure}[h!]
\centering
\includegraphics[width=0.485\textwidth]{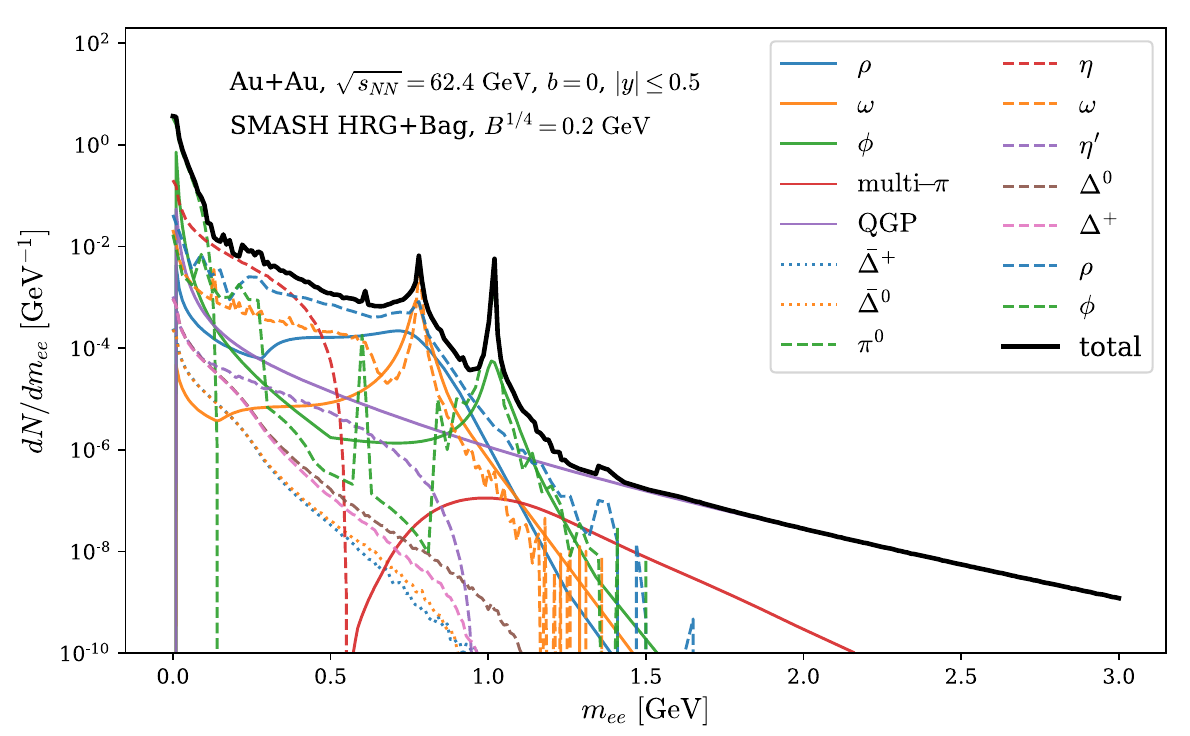}%
\hspace{0.01\textwidth}
\includegraphics[width=0.48\textwidth]{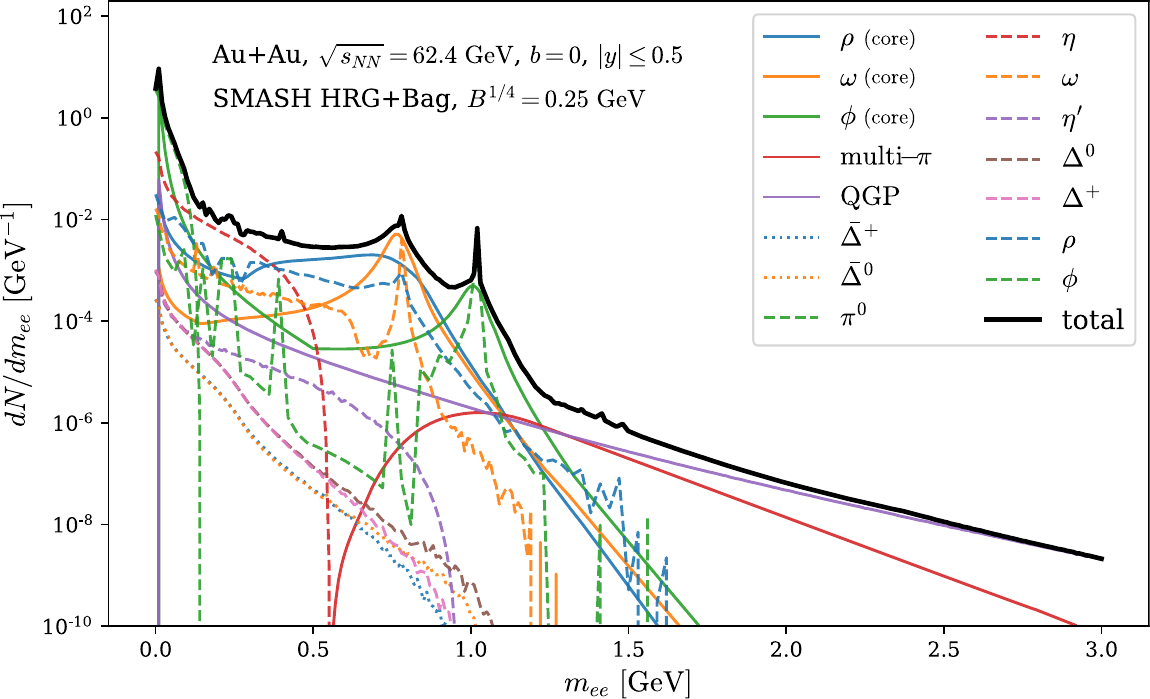}
\caption{Left (Right): dielectron spectra using the Bag+HRG EoS for a weak (strong) bag. Solid (Dashed and dotted) lines indicate contributions from the fluid (hadrons).}
\label{fig:Bag_spectra}
\end{figure}

\begin{figure}[h!]
    \centering
    \includegraphics[width=0.325\textwidth]{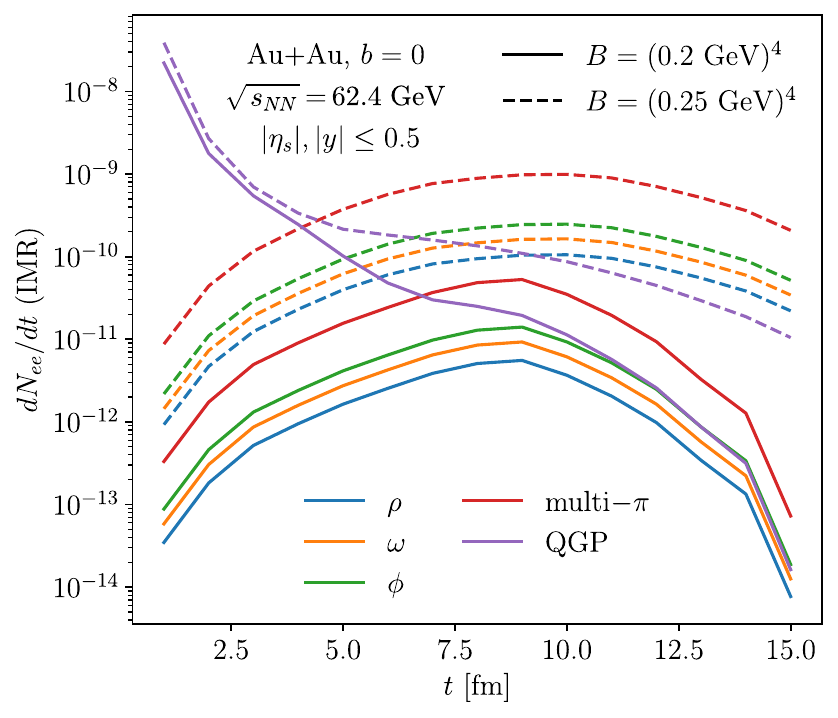}%
    \hspace{0.01\textwidth}%
    \includegraphics[width=0.66\linewidth]{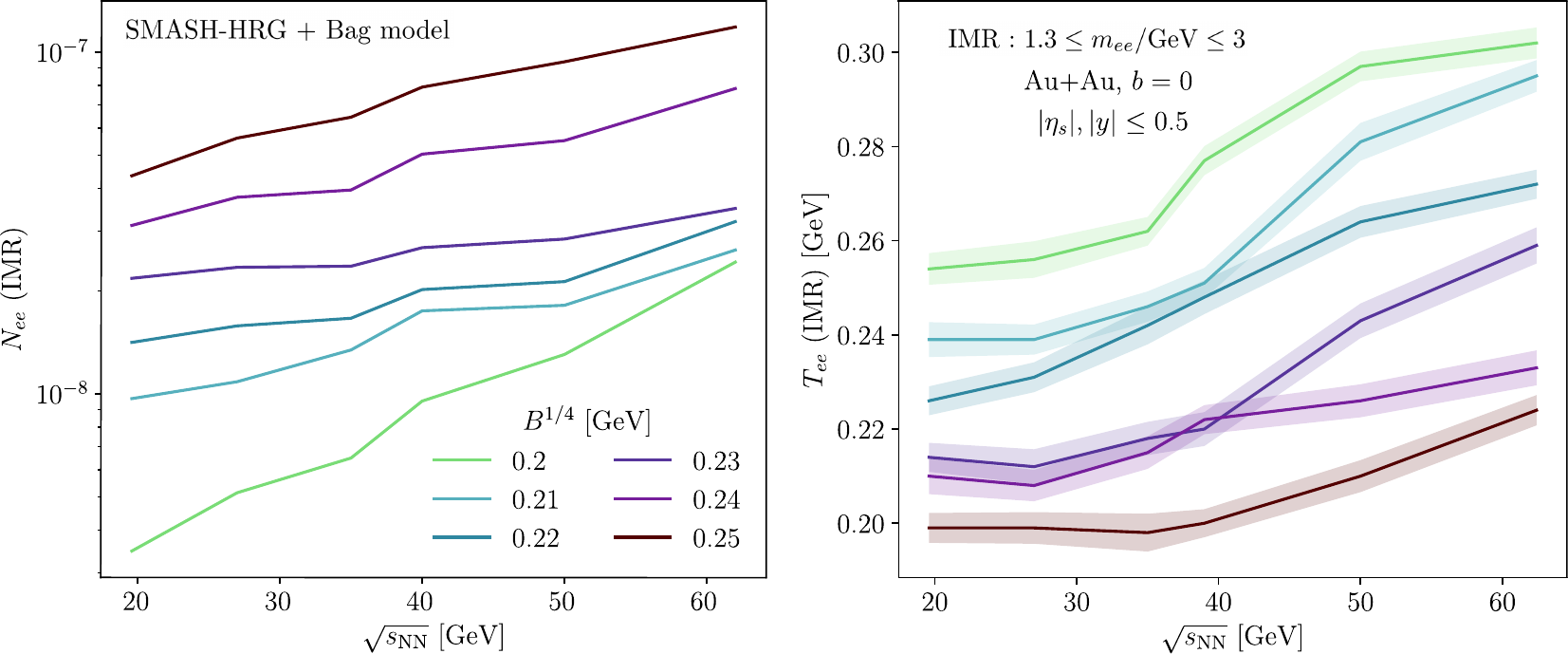}
    \caption{Left: Time dependence of each source in the IMR from the fluid, for different bag parameters. Middle (Right): Excitation function of yield (effective temperature). Error bands represent fit uncertainty.}
    \label{fig:Bag_evolution_excitation}
\end{figure}

Figure~\ref{fig:Bag_spectra} shows the contributing sources to the dielectron spectra of central Au+Au collisions at $\sqrtsNN=62.4\ \GeV$, using the EoS \eqref{eq:Bag}. The left plot shows a ``weak'' bag, with $B^{1/4}=0.2\ \GeV$, and the right a ``strong'' bag, with $B^{1/4}=0.25\ \GeV$. The low invariant mass range ($m_{ee}\leq m_\phi$) displays a competition between core and corona contributions, which reflects how much time the system spends in each stage. The weak bag has a larger corona contribution, because the smaller latent heat means that the fluid stage is shorter. Moreover, the weak bag has narrower peaks around the pole masses of vector mesons, since the transition to HRG starts at lower temperatures, leading to less broadened spectral functions. In the IMR, the QGP rate dominates, with a subleading contribution from multi-$\pi$ radiation which becomes more important with higher $B$. This is explained by the time evolution of these sources, shown in the leftmost plot of Fig. \ref{fig:Bag_evolution_excitation}: initially, the system is fully in the QGP phase. As time passes and the fluid cools down, it transitions to the HRG. In the strong bag, this happens earlier and the curves are less sloped, indicating that the transition lasts longer.

The middle and rightmost plots of Fig.~\ref{fig:Bag_evolution_excitation} show respectively the excitation function of the yield and effective temperature of IMR dielectrons for different bag parameters. As mentioned above, a larger $B$ carries with more latent heat and therefore a longer transition, giving the fluid more time to radiate. However, since the multi-$\pi$ term is relatively more important in stronger bags and is associated with a softer spectrum (cf. Fig.~\ref{fig:Bag_spectra}), it reduces the inverse slope $T_\mathrm{eff}$.

\mysection{Ising 2D T'-Expansion Scheme.} A realistic EoS is needed to make quantitative statements about the effects of a phase transition. For this, we use the Ising-2DTExS module of the MUSES collaboration \cite{Kahangirwe:2024cny}. It is constructed by mapping thermodynamic quantities from the 3D Ising model onto QCD variables, as the two theories belong to the same universality class. In this section, we use the shipped configuration with a CEP at $(T^c,\mu_B^c)=(0.139,0.4)\ \GeV$. Furthermore, the module allows for changing these values along the transition line, which we exploit by setting it to the maximum $\mu_B$ value, effectively mimicking a crossover. The two settings differ only near the transition region -- the confined and deconfined regions are the same --, so any difference in observables arises exclusively from the system passing through a first-order transition, unlike in \cite{Seck:2020qbx,Savchuk:2022aev} or the Bag+HRG EoS.

\begin{figure}[h!]
\centering
\includegraphics[height=13.5em]{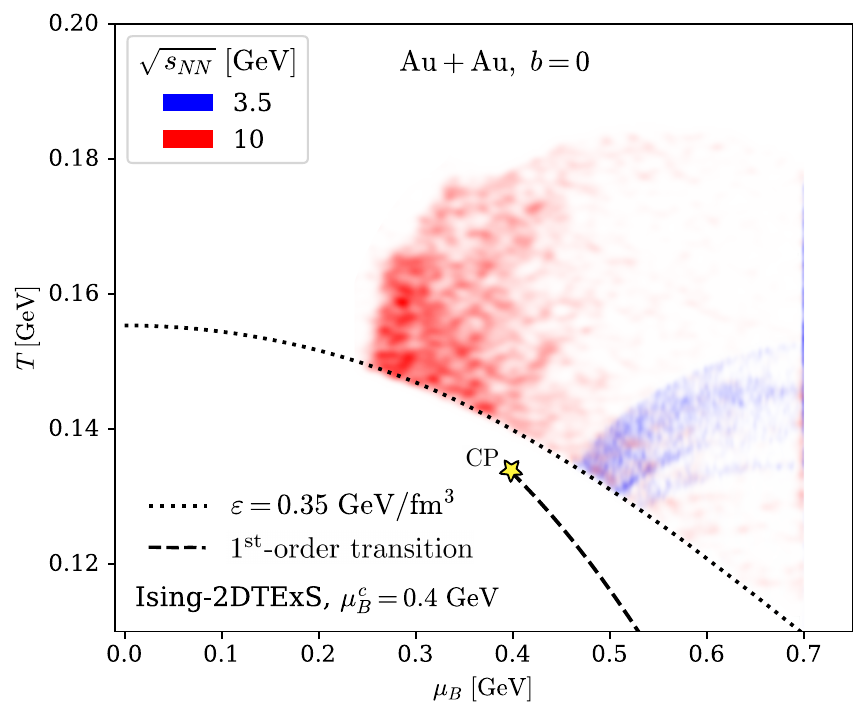}%
\includegraphics[height=13.5em]{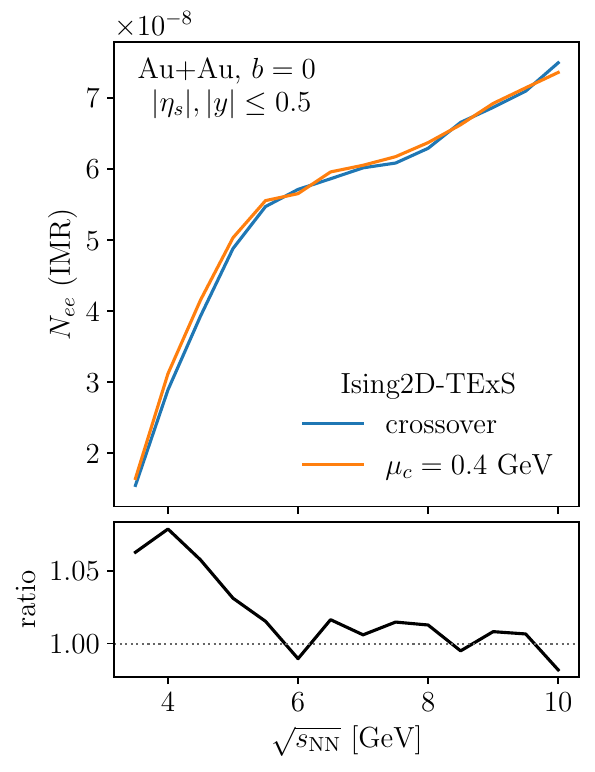}%
\hspace{0.005\textwidth}
\includegraphics[height=13em]{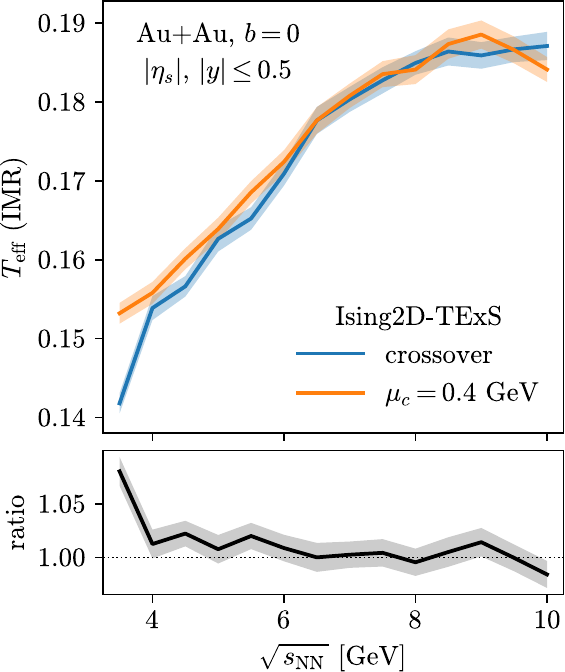}
\caption{Left: Distribution of initial fluid cells in the phase diagram using the Ising-2DTExS EoS. Middle (Right): Excitation function of yield (effective temperature). Error bands represent fit uncertainty.}
\label{fig:Ising_results}
\end{figure}

The left plot of Fig.~\ref{fig:Ising_results} shows the distribution of cells in the phase diagram when the fluid is initially formed. We see that a collision at $\sqrtsNN=3.5\ \GeV$ passes completely through the first-order transition, while a $10\ \GeV$ collision starts mostly above it, in the crossover region. As this falls within the low-to-intermediate beam energy range, we use DynFlu initial conditions, such that all cells start above the threshold $\varepsilon_\mathrm{th}=0.35\ \GeV/\mathrm{fm}^3$. Moreover, this plot also shows a drawback of this module: it is limited up to $\mu_B=0.7\ \GeV$, and so does not cover the densest parts of the system. We deal with this by setting the temperature of the above-bound cells to a value that maintains the same energy density. This breaks baryon number conservation, but since it happens in both CEP and crossover EoSs, we assume it does not affect the comparison significantly. 

The middle and right plots of Fig.~\ref{fig:Ising_results} show the excitation function of IMR yield and effective temperature for both EoSs. As expected, both quantities are consistent with a monotonic increase, which is more rapid at lower beam energies. For $\sqrtsNN\leq 5.5\ \GeV$, the ratio between the curves with/without a CEP displays a consistent enhancement, revealing that the yield increases in the presence of a phase transition by at most 7\%. The effective temperature plot also hints at an enhancement, but the ratio is consistent with unity within fit uncertainty.  

\section{Discussion}

We have explored the impact of phase transitions on 
dilepton production in two scenarios in the SMASH-vHLLE hybrid model. The qualitative analysis with a Bag+HRG EoS shows a nontrivial model dependence of the IMR yields and effective temperature $T_\mathrm{eff}$ in both dynamical evolution and equation of state; in particular the interplay between different dilepton sources causes $T_\mathrm{eff}$ to decrease in bags with a larger latent heat. With a realistic EoS based on the Ising universality class, we found a non-negligible enhancement of (at most) 7\% in the IMR yield when the system passes through a first order phase transition. While small, this effect is assuredly caused by the transition itself, and not due to a different EoS coverage in $(T,\mu_B)$. However, we highlight that the present model does not include critical phenomena or spinodal decomposition, which likely change this value. We make the further caveat that the derivation of thermal rates in \cite{Rapp:1999us,vanHees:2007th} assumes a medium in equilibrium, which is not fully consistent with viscous hydrodynamics. Nonetheless, the results shown here are consistent with the expectation of an observable enhancement in dilepton production in the presence of a phase transition.

\section*{Acknowledgements}

 We acknowledge the support by the Deutsche Forschungsgemeinschaft (DFG, German Research Foundation) – Project number 315477589 – TRR 211. Computational resources have been provided by the GreenCube at GSI.

\bibliographystyle{apsrev4-1}
\bibliography{References}

\end{document}